\documentclass[%
 reprint,
superscriptaddress,
 amsmath,amssymb,
 aps,
]{revtex4-1}

\usepackage{graphicx}
\usepackage{dcolumn}
\usepackage{bm}
\usepackage{qcircuit}
\usepackage{physics}
\usepackage{mathtools}
\usepackage{xcolor}
\usepackage{booktabs}
\usepackage{multirow}

\usepackage[utf8]{inputenc}

\begin{document}

\preprint{APS/123-QED}

\title{Optimized noise-assisted 
simulation of the Lindblad equation with time-dependent coefficients on a noisy quantum processor}

\author{José D. Guimarães}

\affiliation{%
Institute of Theoretical Physics and IQST, Ulm University, Albert-Einstein-Allee 11 89081, Ulm, Germany.
}%

\affiliation{%
Centro de Física das Universidades do Minho e do Porto, Braga 4710-057, Portugal
}%

\affiliation{%
Intl. Iberian Nanotechnology Laboratory,
Av. Mestre Jos{\'e} Veiga s/n, Braga 4715-330, Portugal.
}%

\author{Antonio Ruiz-Molero}

\affiliation{%
Intl. Iberian Nanotechnology Laboratory,
Av. Mestre Jos{\'e} Veiga s/n, Braga 4715-330, Portugal.
}%

\affiliation{Departamento de Inform\'atica, Universidade do Minho, Braga 4710-057, Portugal}

\author{James Lim} 

\affiliation{%
 Institute of Theoretical Physics and IQST, Ulm University, Albert-Einstein-Allee 11 89081, Ulm, Germany.
}%
\author{Mikhail I. Vasilevskiy}

\affiliation{%
Centro de Física das Universidades do Minho e do Porto, Braga 4710-057, Portugal
}%

\affiliation{%
Intl. Iberian Nanotechnology Laboratory,
Av. Mestre Jos{\'e} Veiga s/n, Braga 4715-330, Portugal.
}%

\author{Susana F. Huelga}

\affiliation{%
Institute of Theoretical Physics and IQST, Ulm University, Albert-Einstein-Allee 11 89081, Ulm, Germany.
}%
\author{Martin B. Plenio}

\affiliation{%
Institute of Theoretical Physics and IQST, Ulm University, Albert-Einstein-Allee 11 89081, Ulm, Germany.
}%

\begin{abstract}
Noise in quantum devices is generally considered detrimental to computational accuracy. However, the recent proposal of noise-assisted simulation has demonstrated that noise can be an asset in digital quantum simulations of open systems on Noisy Intermediate-Scale Quantum (NISQ) devices. In this context, we introduce an optimized decoherence rate control scheme that can significantly reduce computational requirements by multiple orders of magnitude, in comparison to the original noise-assisted simulation. We further extend this approach to encompass Lindblad equations with time-dependent coefficients, using only quantum error characterization and mitigation techniques. This extension allows for the perturbative simulation of non-Markovian dynamics on NISQ devices, eliminating the need for ancilla qubits or mid-circuit measurements. Our contributions are validated through numerical experiments on an emulated IBMQ device. Overall, our work offers valuable optimizations that bring current quantum processors closer to effectively simulating realistic open systems.
\end{abstract}

\maketitle


\section{Introduction}
Open quantum systems are omnipresent in the realm of quantum mechanics, as the ideal isolation of a quantum system from its environment is practically unattainable. Consequently, elucidating the dynamical behavior and inherent properties of these systems is paramount for both a profound understanding of natural quantum phenomena beyond the realm of highly controlled system-environment interactions \cite{chin2013role,caycedo2022exact,cao2020quantum} and the advancement of efficient quantum technologies \cite{somoza2023driving,bravyi2022future,wang2020integrated}. Conventional simulation methodologies for these complex systems often encounter computational bottlenecks when deployed on classical computers due to the exponential surge in required computational resources. Likewise, quantum computers, especially in their current Noisy Intermediate-Scale Quantum (NISQ) phase, grapple with challenges rooted in the absence of fault tolerance.

The task of simulating open quantum systems is further complicated by the exceedingly large number of degrees of freedom of the environment which may not be eliminated in the presence of memory effects \cite{rivasQuantumNonMarkovianityCharacterization2014}. Among the array of methods aimed at addressing these challenges on classical \cite{nusseler2022fingerprint,tamascelli2019efficient,somoza2019dissipation,tamascelli2018nonperturbative,gribben2022exact,tanimura2020numerically,weimer2021simulation,mohseni2014quantum} and quantum platforms—both analog \cite{mostame2012quantum,kim2022analog,lemmer2018trapped,gorman2018engineering,daley2022practical,polla2021quantum} and digital \cite{guimaraes2023noise,barreiro2011open,lemmer2018trapped,cleve2016efficient,wang2011quantum,wang2023simulating,kamakari2022digital,schlimgen2021quantum,georgescu2014quantum,miessen2023quantum,nielsen2002quantum}—the noise-assisted simulation technique on quantum processors that was introduced in \cite{guimaraes2023noise} emerges as a viable option for NISQ devices. This method leverages quantum error characterization and mitigation techniques to simulate Lindblad equations with time-independent coefficients, effectively turning the intrinsic noise of NISQ devices from a drawback into a computational utility.

In this work, we explore two improvements designed to enhance the computational efficiency of the noise-assisted simulation method. Firstly, we incorporate the principle of locality constraints in error mitigation \cite{tran2023locality} to the existing simulation framework, aimed at reducing the sampling cost by optimizing error mitigation procedures. The second consists of a new decoherence rate control scheme, designed to dynamically adjust error rates during the simulation, resulting in lower sampling costs. Notably, our optimizations have the potential to reduce the computational resources required by orders of magnitude when compared to the original noise-assisted simulation approach \cite{guimaraes2023noise}. Additionally, we extend the methodology of \cite{guimaraes2023noise} to enable the simulation of the Lindblad equation with time-dependent coefficients, thereby allowing for the exploration of non-Markovian dynamics.

The paper is organized as follows: Section~\ref{time_evol_sec} reviews the foundational time-evolution simulation and error characterization methods from the original noise-assisted digital quantum simulation technique~\cite{guimaraes2023noise}. It also explores Probabilistic Error Cancellation~\cite{temme2017error,endo2018practical,sun2021mitigating,van2023probabilistic,cai2022quantum,takagi2022fundamental,suzuki2022quantum,strikis2021learning,guo2022quantum,piveteau2022quasiprobability},  focusing on its localized error mitigation variant as proposed in \cite{tran2023locality} and its partial and layer-dependent mitigation form. Section~\ref{sec: decoherence_schemes} introduces our decoherence rate control scheme and assesses its performance relative to the scheme used in the original noise-assisted simulation \cite{guimaraes2023noise}.  Section~\ref{sec: simulation_time_dependent_theory} discusses our extension of the technique to simulate Lindblad equations with time-dependent coefficients. Finally, Section~\ref{sec: num_impl} offers empirical validations of our optimized noise-assisted simulation technique, featuring case studies that capture non-Markovian behavior in open quantum systems.

\section{Encoding of the time evolution in a noisy quantum circuit} \label{time_evol_sec}

In quantum computation, the Trotter-Suzuki product formula serves as a standard technique for approximating the time evolution of closed quantum systems \cite{childs2021theory,csahinouglu2021hamiltonian,clinton2021hamiltonian,childs2018toward}. However, its implementation on real-world quantum processors introduces an inherent level of noise due to gate imperfections. This noise can be modeled accurately by Lindblad dynamics~\cite{van2023probabilistic,ferracin2022efficiently,erhard2019characterizing,chen2023learnability}, a framework that underpins the noise-assisted simulation approach discussed in Ref.~\cite{guimaraes2023noise}. In this section, we review the encoding of time-evolution in a noisy quantum circuit as introduced in Ref.~\cite{guimaraes2023noise}. In particular, we delve into the specific mechanics of implementing the Trotter-Suzuki product formula in a quantum processor and elucidate how this leads to noise dynamics effectively described by the Lindblad equation.

\subsection{Trotter-Suzuki product formula}
 We consider a Hamiltonian decomposed into a sum of $N$ tensor products of Pauli matrices $\hat{X},\hat{Y},\hat{Z}$ acting on the quantum system encoded in the qubits,
\begin{equation}
    \hat{H} = \sum_{j=1}^{N}\hat{H}_{j},\quad \hat{H}_{j}=\beta_{j} \hat{P}_{j}, \label{ham_pauli} 
\end{equation}
where $\beta_{j} \in \mathbb{R}$ and $\hat{P}_{j}=\{\hat{X},\hat{Y},\hat{Z}\}^{\otimes n_j}$ is a Pauli string acting on $n_j$ qubits.

In this work, we focus on the first-order Trotter-Suzuki product formula as the method to simulate the time evolution of a quantum system over time $t$. Such formula can be expressed as
\begin{equation}
    e^{-i\hat{H}t} \approx \left[\prod_{j=1}^{N} e^{-i \hat{H}_{j}\Delta t}\right]^{D}, \label{trotter_equation}
\end{equation}
where $\Delta t$ is a finite Trotter time-step. Higher orders of the Trotter-Suzuki product formula may also be employed \cite{guimaraes2023noise}. The total time evolution is then described by $D = t/\Delta t$ Trotter layers.

On current noisy devices, the implementation of each quantum gate in the Trotter layers, e.g. $e^{-i\hat{H}_j \Delta t}$ in Eq.~\eqref{trotter_equation}, is faulty, thus introduces noise in the system. The time evolution of a quantum system encoded in the qubits depends thus on both the Hamiltonian implemented on a quantum circuit and on the parameters that characterize the intrinsic noise of the quantum device. As shown in previous studies on superconducting qubits~\cite{van2023probabilistic,ferracin2022efficiently,guimaraes2023noise}, the evolution of the density matrix $\hat{\rho}(t)$ of the open system encoded in the qubits can be accurately described by a Markovian quantum master equation expressed as,
\begin{equation}
    \frac{d\hat{\rho}(t)}{dt} = \mathcal{L}[\hat{\rho}(t)] = -i[\hat{H},\hat{\rho}(t)]+\mathcal{D}_{\rm QC}[\hat{\rho}(t)], \label{Lindblad_eq}
\end{equation}
where $\mathcal{D}_{\rm QC}[\hat{\rho}(t)]$ represents a Lindblad dissipator which describes the intrinsic noise of the quantum circuit implemented on the noisy quantum device.

As shown in \cite{guimaraes2023noise}, $\mathcal{D}_{\rm QC}[\hat{\rho}(t)]$ can be characterized and adjusted to simulate a target Lindblad equation with time-independent decoherence rates. In order to achieve this, one must first characterize the noise acting in the quantum circuit. Through the use of Cycle Benchmarking \cite{erhard2019characterizing}, a noise benchmarking technique, the sparse stochastic Pauli noise acting on the qubits over one Trotter layer can be learned \cite{flammia2020efficient,van2023probabilistic,chen2023learnability}. Assuming that the creation of long-ranged spatially correlated noise in one Trotter layer is negligible (as recently demonstrated in superconducting quantum platforms \cite{guimaraes2023noise,van2023probabilistic,hashim2020randomized}), the noise channel acting over one Trotter layer can be approximated as follows \cite{van2023probabilistic},
\begin{align}
    \mathcal{E}(\hat{\rho}) & =\prod_{k}\left(w_k \hat{I}_k \hat{\rho} \hat{I}_k + (1-w_k)\hat{P}_{k}\hat{\rho}\hat{P}_{k}\right),\label{stochastic_channel}\\
    w_{k} & = (1+e^{-2\epsilon_{k}})/2\label{definition_wk}
\end{align}
where $\hat{P}_{k}$ is a 2-qubit Pauli string acting on nearest-neighbour qubits in the quantum circuit with an associated error probability $\epsilon_{k}$, which may be characterized via Cycle Benchmarking \cite{erhard2019characterizing,flammia2020efficient}. Here, we assume that only nearest-neighbor interactions between qubits are possible, such as in current superconducting quantum devices. In order to transform all noise in one Trotter layer into stochastic Pauli channels such as the one in Eq.~\eqref{stochastic_channel}, Randomized Compiling (RC) is applied to all (non-negligible) noisy operations (see Refs. \cite{erhard2019characterizing,hashim2020randomized,guimaraes2023noise} for more details about this error transformation technique). Following Ref.~\cite{guimaraes2023noise}, the implementation of RC is performed on the noise characterization and quantum simulation circuits, usually to all 2-qubit gates since these are the noisiest operations in a quantum circuit \cite{van2023probabilistic,hashim2020randomized}. 
By applying RC to the (non-negligible) noisy operations, the transformed noise in the quantum circuit is approximately given by Eq.~\eqref{stochastic_channel} and, when evolving the qubits via the Trotter-Suzuki product formula, the following Lindblad dissipator is simulated in the quantum circuit,
\begin{align}
    \mathcal{D}_{\rm QC}[\hat{\rho}] &= \sum_{k}\gamma_k \left(\hat{P}_k\hat{\rho}\hat{P}_k - \hat{\rho}\right),\label{eq:D_stochastic_controlled}\\
    \gamma_k &= \epsilon_k/\Delta t.\label{eq:gamma_stochastic_uncontrolled}
\end{align}
Here the decoherence rates, $\gamma_k$, are associated with the 2-qubit Pauli strings, $\hat{P}_k$, and are defined as functions of the characterized error probabilities, $\epsilon_k$, and the chosen Trotter time-step, $\Delta t$. 

As demonstrated in Ref.~\cite{guimaraes2023noise}, the decoherence rates $\gamma_{k}$ can be selectively increased or decreased. The former, by reducing $\Delta t$ and the latter, by implementing a quantum error mitigation technique, the Probabilistic Error Cancellation \cite{temme2017error,endo2018practical,sun2021mitigating,van2023probabilistic,cai2022quantum,takagi2022fundamental,suzuki2022quantum,strikis2021learning,guo2022quantum,piveteau2022quasiprobability} that decreases the error probability ($\epsilon_{k}$) of a stochastic Pauli channel $k$. These two protocol steps allow the decoherence rates $\gamma_{k}$ to be controlled according to Eq.~\eqref{eq:gamma_stochastic_uncontrolled}. 

\subsection{Probabilistic Error Cancellation}\label{sec: PEC}

In this section, we start by describing the conventional PEC technique~\cite{temme2017error,endo2018practical,sun2021mitigating,van2023probabilistic,cai2022quantum,takagi2022fundamental,suzuki2022quantum,strikis2021learning,guo2022quantum,piveteau2022quasiprobability}, used to fully mitigate the noise in a quantum circuit. Then, we briefly introduce partial error mitigation (see Ref.~\cite{guimaraes2023noise} for more details) and we discuss circuit-aware error mitigation \cite{tran2023locality}. We find that  the latter requires substantially less measurements for local observables than the traditional circuit-blind mitigation protocol as illustrated in Fig.~\ref{fig:local_mitigation}. Lastly, we discuss the layer-dependent error mitigation with PEC, which allows for the control of time-dependent noise in the context of its application to Trotter-type circuits.

\subsubsection{Full error mitigation}

PEC starts by identifying the noise channel, $\mathcal{E}$, acting on 
an ideal, noiseless circuit $\mathcal{C}(\hat{\rho})=\hat{U}_{k}(\Delta t)\hat{\rho}\hat{U}_{k}^{\dagger}(\Delta t)$, for instance, a Trotter layer describing the Hamiltonian dynamics of a quantum system over a Trotter time-step $\Delta t$. As discussed in Section~\ref{time_evol_sec}, one can characterize the noise of a quantum circuit by $2$-qubit stochastic Pauli noise channels $\mathcal{E}$ (see Eq.~\eqref{stochastic_channel}). To fully mitigate the characterized noise, the conventional PEC has considered the inverted noise channel, $\mathcal{E}^{-1}$, applied to the noisy quantum circuit, namely implemented after $\mathcal{U}(\hat{\rho}) = \mathcal{E}\cdot\mathcal{C}(\hat{\rho})$. This is illustrated in Fig.~\ref{fig:local_mitigation}(a). An inverted stochastic Pauli channel, $\mathcal{E}^{-1}$, can be exactly formulated as \cite{van2023probabilistic},
\begin{align}  
\mathcal{E}^{-1}(\hat{\rho}) & = C_{mit}\prod_{k}\left(w_k \hat{I}_k \hat{\rho} \hat{I}_k - (1-w_k)\hat{P}_{k}\hat{\rho}\hat{P}_{k}\right),\label{inverted_stochastic_channel}\\
C_{mit} & = \exp{2\sum_{k}\epsilon_{k}}.\label{mitigationcost}
\end{align}
Since $\mathcal{E}^{-1}$ is not a complete-positive (CP) map, one cannot implement it directly in the quantum circuit. 
The application of the non-CP map $\mathcal{E}^{-1}$ is done in a probabilistic fashion \cite{endo2018practical}, where one of the Pauli operators $\hat{P}_{k}$ is randomly sampled based on the probabilities $w_{k}$ and applied in a Trotter layer. On a quantum circuit with $D=t/\Delta t$ Trotter layers, the probabilistic non-CP map is applied $D$ times with Pauli operators $\hat{P}_{k}$ independently sampled for each Trotter layer. For a circuit with $n$ qubits, the inverted noise channel, $\mathcal{E}^{-1}$, is applied to each pair of nearest-neighbour qubits. We note however that multiple cancellations of single-qubit Pauli noise channels can occur on adjacent qubits, thus the probabilities $w_{k}$ must be changed accordingly \cite{guimaraes2023noise}.

In PEC, the outcome of an observable $\hat{O}$ measured in a noise-mitigated quantum circuit is multiplied by the mitigation cost $C_{\rm mit}$ and other prefactors, namely, $\langle \hat{O}\rangle_{\rm mit} \approx \overline{\langle \hat{O}\rangle}\prod_{d=1}^{D}\prod_{m}C^{(m)}_{\rm mit}$, where $m$ describes different subgroups (nearest-neighbour pairs) of qubits under the action of the $2$-qubit Pauli noise channels. For a noise-mitigated density matrix $\hat{\rho}_{\rm mit}(t)$, the (not normalized) expectation value $\text{Tr}[\hat{O}\hat{\rho}_{\rm mit}(t)]$ of the observable $\hat{O}$ is obtained as $\overline{\langle \hat{O}\rangle}$ by classically averaging the outcomes of the PEC scheme (multiplied by a phase factor $\pm 1$ \cite{endo2018practical}), requiring multiple copies of quantum circuits. The total mitigation cost of the quantum simulation is then defined as
\begin{equation}
    C_{\rm tot}=\prod_{d=1}^{D}\prod_{m}C^{(m)}_{\rm mit},\label{cost_definition}
\end{equation}
 and it normalizes the observable measurement $\overline{\langle \hat{O}\rangle}$ to its average value. As shown in previous works \cite{endo2018practical,guimaraes2023noise,cai2022quantum}, the mitigation cost can be analytically estimated as
\begin{equation}
    C_{\rm tot} \sim e^{\lambda (n-1) D \overline{\epsilon}}, \label{cost_scaling2}
\end{equation}
where $\lambda$ is a parameter to be fitted by experiments (typically $\lambda<1$ \cite{guimaraes2023noise}) and $\overline{\epsilon}=\sum_{1}^{15}\overline{\epsilon_{k}}$ is the averaged total error probability over all pairs of nearest-neighbour qubits. Since the variance of a measured observable with PEC scales as $\propto C_{\rm tot}^2$ \cite{endo2018practical}, the implementation of full error mitigation on a quantum circuit with a large circuit volume $V=nD$ is practically unfeasible due to the exponentially increasing number of quantum circuits (see Eq.~\eqref{cost_scaling2}) that are needed to be executed on the quantum processor to accurately estimate an observable, i.e. with low statistical error.

\subsubsection{Partial error mitigation}

On the other hand, Partial Probabilistic Error Cancellation may be implemented to leverage the noise of the quantum circuits in view of simulating open systems in a quantum processor \cite{guimaraes2023noise}. This is, the noise is \emph{only partially} and \emph{selectively} mitigated, such that the probabilities $w_{k}$ in Eq.~\eqref{definition_wk} and the mitigation cost $C_{\rm mit}$ in Eq.~\eqref{mitigationcost} are renormalized, i.e. the total error probability $\epsilon_{k}$ is replaced by the (generally smaller) partially mitigated error probability $\epsilon'_{k}=r_{k}\epsilon_{k},$ with $ r_{k} \in [0,1]$ denoting a mitigation factor associated to a stochastic Pauli channel $k$. This adjustment allows Probabilistic Error Cancellation to be used for quantum simulations of open systems with a reduced total mitigation cost, where $\overline{\epsilon}$ in Eq.~\eqref{cost_scaling2} is replaced by the averaged total mitigated error probability $\overline{\epsilon_{r}}=\sum_{k=1}^{15}\overline{\epsilon'_{k}}=\sum_{k=1}^{15}\overline{r_{k}\epsilon_{k}}$. Since $r_{k} \in [0,1]$, the total mitigation cost in Eq.~\eqref{cost_scaling2} may be exponentially reduced, hence allowing open system simulations with a higher circuit volume to be executed on a noisy quantum processor \cite{guimaraes2023noise}.

\subsubsection{Local error mitigation}

\begin{figure}
\centering
\includegraphics[width=0.48\textwidth]{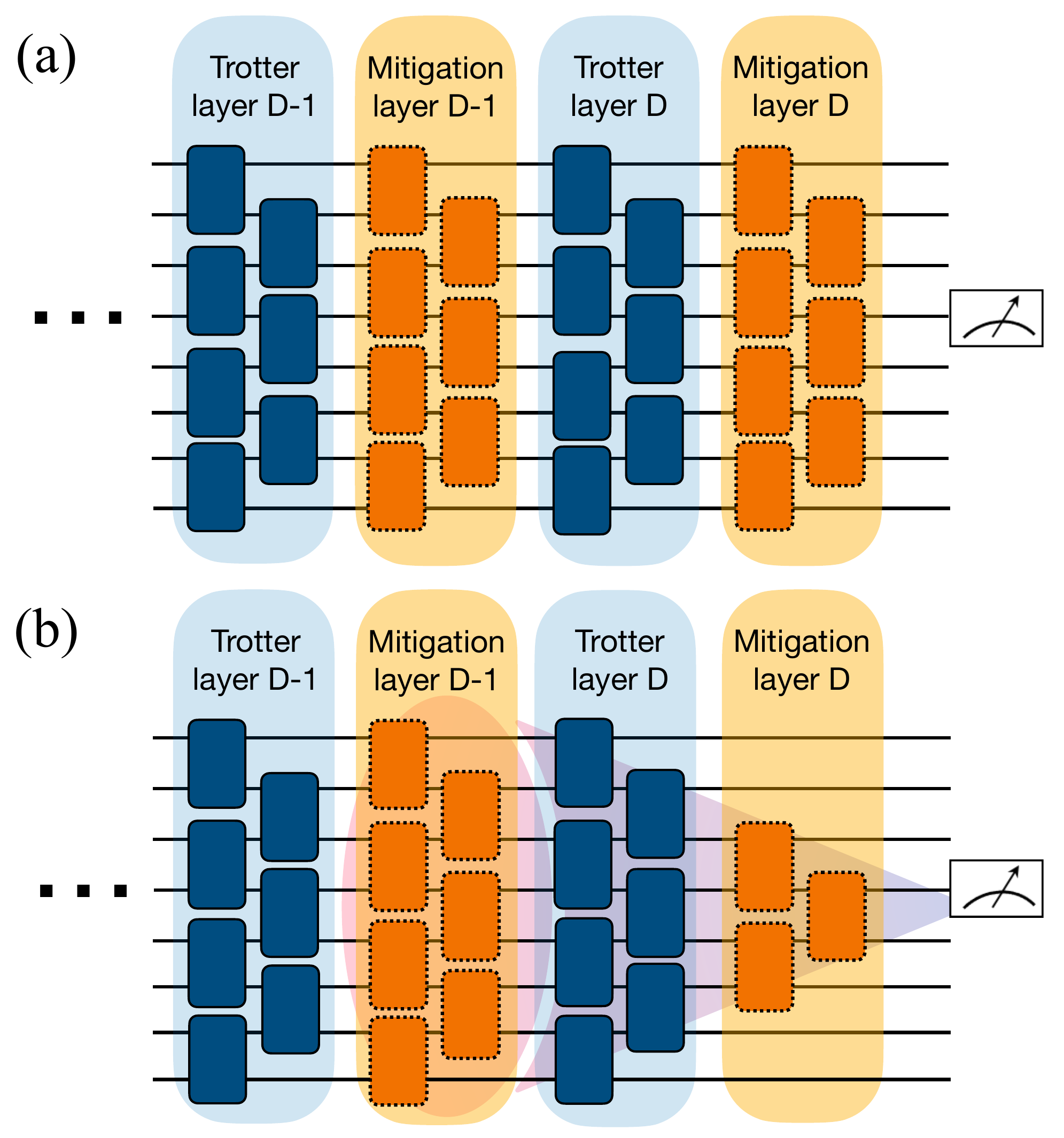}
\caption{Circuit-blind and circuit-aware mitigation protocols. Blue (orange) blocks denote noisy 2-qubit and 1-qubit gates (sampled Pauli strings used for noise mitigation). (a) PEC applied to each Trotter layer and blind to the structure of the quantum circuit. (b) Circuit-aware PEC enables one to take advantage of the locality of the circuit, thus a light-cone mitigation is sufficient to obtain accurate local-measurement outcomes.}
\label{fig:local_mitigation}
\end{figure}

Recently, it has been proposed \cite{tran2023locality} that a reduction of the total mitigation cost is possible when a $k-$local observable is desired to be measured on a quantum circuit. Such models enable a "light-cone" error mitigation over the quantum circuit as shown in Fig.~\ref{fig:local_mitigation}(b), hence reducing the number of stochastic Pauli noise channels that are required to be mitigated relatively to a mitigation protocol blind to the structure of the quantum circuit (Fig.~\ref{fig:local_mitigation}(a)). For instance, the scaling of the total mitigation cost of a light-cone partial error mitigation for an observable acting on $k=1$ or $k=2$ nearest-neighbour qubits can be derived directly from the structure of the 1D quantum circuit shown in Fig.~\ref{fig:local_mitigation}. A straightforward estimate of the total mitigation cost for the circuit in Fig.~\ref{fig:local_mitigation} can be given as follows,
\begin{align}
C_{\rm tot}^{\rm (loc)}&=\prod_{d=1}^{D}C^{\rm (loc)}_{\rm iter}(d), \label{local_mitigation_cost}  \\
C^{\rm (loc)}_{\rm iter} (d)  & \sim
\begin{cases}
        \exp{\lambda (1+2k+2d)\overline{\epsilon_{r}}}, \quad 2(1+k+d) < n \\
    \exp{\lambda (n-1) \overline{\epsilon_{r}}}, \quad 2(1+k+d) \geq n,
\end{cases} \label{iteration_mit_cost_local}
\end{align}
where $C^{\rm (loc)}_{\rm iter}(d)$ is the iteration mitigation cost, i.e. the mitigation cost for the $d-$th Trotter layer in the circuit.
A circuit-aware mitigation yields a total mitigation cost (Eq.~\eqref{iteration_mit_cost_local}) that is independent of the number $n$ of qubits for small $D$, in contrast to Eq.~\eqref{cost_scaling2}. To better illustrate the advantages of this optimization, consider a quantum circuit with $n=25$ qubits, $k=1$-qubit observable, $\lambda=0.5$, a total average 2-qubit mitigated error of $\overline{\epsilon_{r}}=0.06$ and a total of $D=15$ Trotter layers. In this case, the number of circuits required to simulate the open system with a circuit-aware mitigation (Fig.~\ref{fig:local_mitigation}(b)) is $3$ orders of magnitude lower than with a circuit-blind mitigation (which is defined by the second case of Eq.~\eqref{iteration_mit_cost_local} and shown in Fig.~\ref{fig:local_mitigation}(a)). Therefore, a circuit-aware mitigation can potentially save orders of magnitude of computational resources for large-scale quantum simulations of open systems in NISQ devices relatively to the traditional circuit-blind error mitigation strategy for local observables.

\subsubsection{Layer-dependent error mitigation}\label{sec: timedependent_PEC}

Partial PEC can also be applied as layer-dependent error mitigation. This is, for each Trotter layer, one may choose different mitigation factors. Consider, for simplicity, a target time-dependent stochastic Lindblad dissipator that one desires to simulate, expressed in terms of Pauli strings,
    \begin{align}
    \mathcal{D}^{\rm (target)}_{\rm QC}(t)[\hat{\rho}] &= \sum_{k}\Gamma_{k}(t) \left(\hat{P}_{k}\hat{\rho}\hat{P}_{k} - \hat{\rho}\right),\label{eq:D_stochastic_nonMarkovian}\\
    \Gamma_{k}(t) &= \epsilon_{k}(t)/\Delta t\label{eq:gamma_stochastic_nonMarkovian}.
\end{align}
where $\Gamma_{k}(t)$ and $\epsilon_{k}(t)$ are defined as functions of time for each Pauli string $k$. Such time-dependent error probabilities may be attained in a noise-assisted digital quantum simulation by assigning to each Trotter layer, i.e. to each interval of time $[t,t+\Delta t]$, a discrete target decoherence rate $\Gamma_{k}(t=D \Delta t)$. We assume here smooth target decoherence rates $\Gamma_{k}(t)$, such that the latter can be accurately discretized. Hence, partial PEC can be used with different sampling probabilities for each Trotter layer. In the context of a Trotter-type time-evolution circuit, this means that the mitigation factor may be chosen to be time-dependent, i.e. different for each Trotter layer. More specifically, for each time $t = D\Delta t$, the target decoherence rates are defined as follows,
\begin{equation}
    \Gamma_{k}[t = D\Delta t]=\frac{\epsilon_{k}[D\Delta t]}{\Delta t}=\frac{\epsilon_{k}(1-r_{k}[D\Delta t])}{\Delta t}. \label{eq: nonMarkovian_discretized_gamma}
\end{equation} 
Therefore, time-dependent decoherence rates in the quantum circuit may be adjusted via layer-dependent partial PEC.

\section{Control of decoherence rates in the quantum circuit}\label{sec: decoherence_schemes}

As previously mentioned, the quantum error mitigation technique Probabilistic Error Cancellation can be used to adjust the error probabilities $\epsilon_{k}$ of a quantum circuit in view of simulating a time-independent Lindblad equation in a quantum processor. Assuming, for simplicity, time-independent decoherence rates, the equation reads:
\begin{equation}
    \mathcal{D}_{\rm target}[\hat{\rho}] = \sum_{k}\Gamma_{k} \left(\hat{P}_k\hat{\rho}\hat{P}_k - \hat{\rho}\right),\label{eq:target_D_stochastic_target}
\end{equation}
where $\Gamma_k$ are the target decoherence rates associated with each 2-qubit Pauli string $\hat{P}_k$. In order to tune the circuit decoherence rates ($\gamma_{k}$) to the desired ones ($\Gamma_k$), a decoherence rate control scheme must be employed. Herein, we discuss two schemes, which we denote by \textit{scheme I} and \textit{scheme II}. We find that the former, proposed in Ref.~\cite{guimaraes2023noise}, is not as resource-efficient as the latter in realistic simulations on NISQ devices (see Sec.~\ref{sec: scheme I}). Therefore, we propose a novel decoherence rate control scheme (II), which reduces computational resources by multiple orders of magnitude relatively to scheme I (see Sec.~\ref{sec: scheme II}). Additionally, we find that scheme II is particularly suited for large-scale realistic simulations of open systems on noisy quantum devices.

\subsection{Scheme I}\label{sec: scheme I}

The tuning of the initial decoherence rates, $\gamma_{k}$, to reach $\Gamma_{k}$ on the noise-assisted quantum simulation may be accomplished in a two-step procedure following Ref.~\cite{guimaraes2023noise}. It consists of \textit{(step 1)} reduction of the Trotter time-step $\Delta t$ to $\Delta t_{max}$ so that larger decoherence rates $\Tilde{\Gamma}_{k}=\epsilon_{k}/\Delta t_{max}>\gamma_{k}$ may be reached as per Eq.~\eqref{eq:gamma_stochastic_uncontrolled} and \textit{(step 2)} selective mitigation of the error probabilities $\epsilon_{k}$ to $(1-r_{k})\epsilon_{k}$ in order to adjust the decoherence rates $\Tilde{\Gamma}_{k}$ to the target ones $\Gamma_{k}$. Using this protocol, the noise-assisted simulation implemented in the quantum processor is described by the Lindblad dissipator in Eq.~\eqref{eq:target_D_stochastic_target} with decoherence rates,
\begin{equation}
        \Gamma_k = (1-r_{k})\epsilon_k/\Delta t_{max},\label{eq:controlled_dec_rates_scheme_I}
\end{equation}
where $r_{k} \in [0,1]$ are mitigation factors that determine the amount of mitigated error $\epsilon'_{k}=r_{k}\epsilon_{k}$ controlled by partial PEC (step 2). $\Delta t_{max}$ is the adjusted Trotter time-step (step 1) defined as
\begin{equation}
    \Delta t_{\rm max} = \epsilon_{\rm k_{\rm max}}/\Gamma_{\rm k_{\rm max}}, \label{delta_t_max}
\end{equation}
where $\{\Gamma_{\rm k_{\rm max}}| k_{\rm max} = \text{argmax}_{k}(\Delta \Gamma_{k})\}$, with $\Delta \Gamma_{k} = \Gamma_{k}-\gamma_{k} > 0$, is the target decoherence rate with the largest positive difference $\Delta \Gamma_{k}$, and $\epsilon_{\rm max}$ is the characterized error probability of the associated stochastic channel $k_{\rm max}$. As shown in Ref.~\cite{guimaraes2023noise}, this protocol enables one to simulate open systems potentially with exponentially lower computational resources than the techniques that require full error mitigation in a noisy quantum device. Even though this protocol can accelerate the deployment of quantum computing to real-world open system problems (see Ref.~\cite{guimaraes2023noise}), we find that this decoherence rate control scheme is not the most efficient to be implemented on current quantum devices. To illustrate this, consider some decoherence rates which are larger than the initial ones, i.e. $\Gamma_{k}>>\gamma_{k}$, whereas others are not, i.e. $\Gamma_{k'\neq k} \lesssim \gamma_{k'}$. Using Eq.~\eqref{eq:controlled_dec_rates_scheme_I} and Eq.~\eqref{delta_t_max} the mitigation factor for each stochastic Pauli channel $k$ is:
\begin{equation}
    r_{k}=1-\frac{\Gamma_{k}\epsilon_{k_{\rm max}}}{\Gamma_{k_{\rm max}}\epsilon_{k}}.\label{mitigation_factor_scheme_I}
\end{equation}
Simply put, for fixed error probabilities $\epsilon_{k}$, the larger the target decoherence rate $\Gamma_{k_{\rm max}}$ is relatively to all others ($\Gamma_{k\neq k_{\rm max}}$), the larger the mitigation factor $r_{k}$ of that stochastic channel ($k$) is. This means that an increase of the target decoherence rate $\Gamma_{k_{\rm max}}$ yields an increase of the mitigation factor of all other channels $k \neq k_{\rm max}$. This global influence of the maximum decoherence rate increase yields an exponentially higher total mitigation cost as per Eq.~\eqref{local_mitigation_cost}, hence the simulation requires a larger number of circuits to be executed. Summarizing, this decoherence rate control scheme I can be employed when the target open system model contains decoherence rates $\Gamma_{k_{\rm max}}\approx \Gamma_{k\neq k_{\rm max}}$. On the other hand, if a broad range of initial or target decoherence rates is present, e.g. such that $\Gamma_{k_{\rm max}} >> \Gamma_{k\neq k_{\rm max}}$, the decoherence rate control scheme I, as introduced in Ref.~\cite{guimaraes2023noise}, is not the most appropriate to be employed. In this case, another decoherence rate control scheme is best suited.
\begin{figure}
\centering
\includegraphics[width=0.3\textwidth]{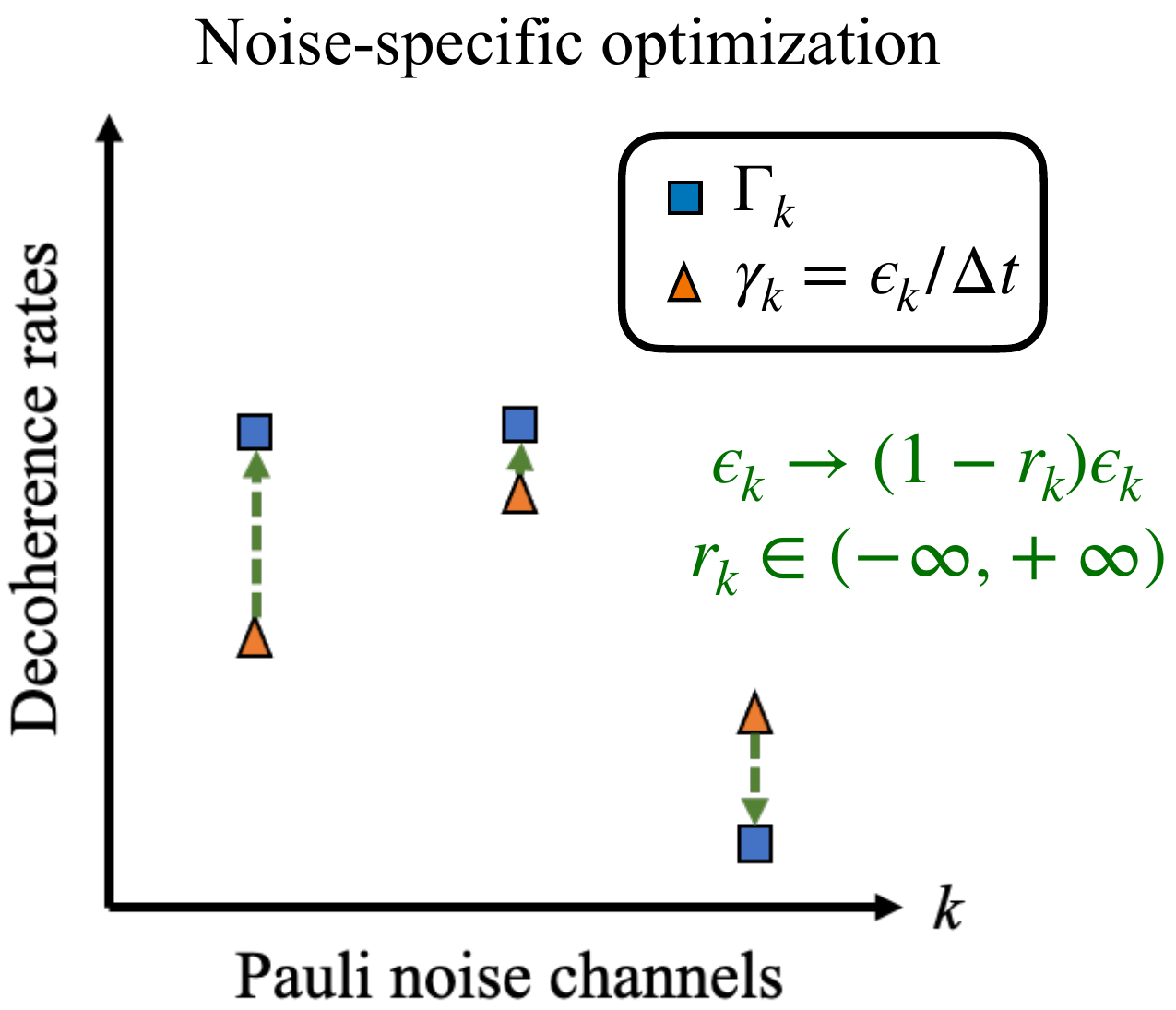}
\caption{Our proposed method to control the decoherence rates. $\gamma_{k}$ ($\Gamma_{k}$) are the initial (target) decoherence rates. In this single-step scheme, some Pauli noise channels $k$ have their decoherence rates increased (noise amplification) by taking the mitigation factor $r_k<0$, whereas others have it decreased (i.e. mitigated) by choosing $r_{k'\neq k}>0$.}
\label{fig:noise_control_scheme}
\end{figure}

\subsection{Scheme II} \label{sec: scheme II}

Herein, we propose a different procedure to reach the target decoherence rates $\Gamma_{k}$ in a quantum circuit. In Fig.~\ref{fig:noise_control_scheme}, we illustrate this scheme. It consists of a one-step procedure, where an arbitrary Trotter time-step $\Delta t$ may be chosen (in contrast to scheme I). The single step consists of a selective reduction (amplification) of the initial decoherence rates $\gamma_{k}$ that are higher (lower) than the target decoherence rates $\Gamma_{k}$. The reduction of noise is performed using partial Probabilistic Error Cancellation and the amplification of noise is done, for instance, via a quantum trajectory approach \cite{chenu2017quantum} or by increasing the noise of the quantum hardware (e.g. as used in the quantum error mitigation technique called Zero-Noise Extrapolation \cite{temme2017error,kim2023evidence,he2020zero,giurgica2020digital,guimaraes2022towards}). The enhancement of stochastic Pauli decoherence rates (and other types of noise) can also be implemented in this scheme using classical noise \cite{chenu2017quantum}, i.e. by implementing stochastic Hamiltonians (for more details, we refer the reader to Ref.~\cite{chenu2017quantum}). The resulting decoherence rates in the quantum circuit are then defined as
\begin{equation}
        \Gamma_k = (1-r_{k})\epsilon_k/\Delta t,\label{eq:controlled_dec_rates_scheme_II}
\end{equation}
where $r_{k} \in (-\infty,+\infty)$. The aforementioned selective noise amplification is expressed in Eq.~\eqref{eq:controlled_dec_rates_scheme_II} by letting the mitigation factor $r_{k}$ have negative values for a particular noise channel $k$ as illustrated in Fig.~\ref{fig:noise_control_scheme}. In this work, we consider the quantum trajectory approach to amplify the noise (the increase of quantum hardware noise will be investigated in a future work), thus in the context of stochastic Pauli channels, we implement a noise amplification map in the quantum circuit, as expressed in Eq.~\eqref{stochastic_channel}, where the probabilities are defined as 
$w_{k}=w_{k}^{\rm (amp)} = (1+e^{2 r_{k}\epsilon_{k}})/2$, where $r_{k}<0$. In each Trotter layer, we apply the respective Pauli operators $\hat{P}_k$ sampled over the probability distribution defined by the amplified error probabilities $w_{k}^{\rm (amp)}$.

\begin{figure*}
\centering
\includegraphics[width=\textwidth]{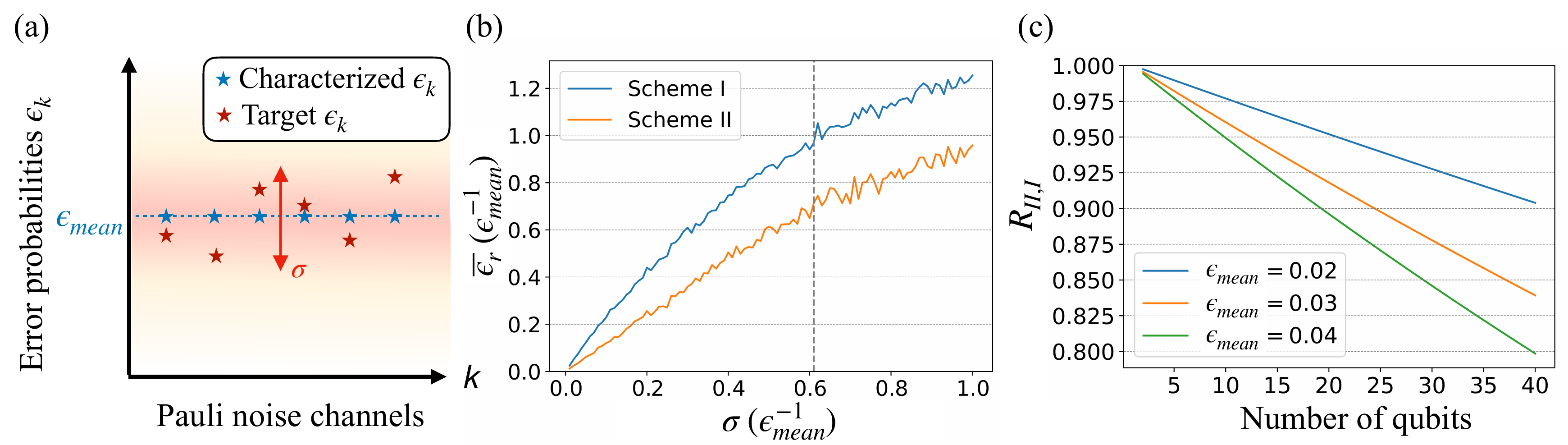}
\caption{Analyses of the relative performance of schemes I and II. (a) We chose the characterized error probabilities in the circuit to be uniform,  $\epsilon_{mean}$ for all stochastic channels. The target error probabilities were sampled from a Gaussian probability distribution with mean $\epsilon_{mean}$ and standard deviation $\sigma \in [\epsilon_{mean}/100,\epsilon_{mean}]$ (we reset negative sampled error probabilities to $0$). (b) Total mitigated error probabilities $\overline{\epsilon_r}$ averaged over $500$ different target error probability samples for each $\sigma$ using schemes I and II. In the former, the Trotter time-step is adjusted, whereas in the later it is not. (c) We calculated the mitigation cost for one Trotter layer using (first case of) Eq.~\eqref{local_mitigation_cost}  for schemes II and I, $C_{\rm iter}^{\rm (II)}$ and $C_{\rm iter}^{\rm (I)}$, respectively, by considering different numbers of qubits and different $\epsilon_{mean}$. We then plot the ratio $R_{II,I}=C_{\rm iter}^{\rm (II)}/C_{\rm iter}^{\rm (I)}$. For this calculation, we considered the $\overline{\epsilon_r}$ that intersect the gray vertical line in (b), where $\sigma = 0.61$. The initial Trotter time-step was chosen to be $\Delta t = 1$ and $\lambda=0.5$ (see Eq.~\eqref{cost_scaling2}).}
\label{fig:scheme_analysis}
\end{figure*}

 Noise reduction, i.e. the partial PEC introduced in Section~\ref{sec: PEC}, is performed to the noise channels $k$ which have positive mitigation factors, whereas noise amplification is applied via Eq.~\eqref{stochastic_channel} with probabilities $w_{k'}^{\rm (amp)}$ to those channels $k' \neq k$ with negative mitigation factors. The sign and magnitude of the mitigation factors is fixed by Eq.~\eqref{eq:controlled_dec_rates_scheme_II}, i.e. by the choice of the target decoherence rates, Trotter time-step and the error probabilities of the quantum device. The sampling procedures required by partial PEC and noise amplification can be done simultaneously. In this proposed decoherence rate control scheme, the mitigation factors $r_k$ are also not required to be upper bounded by 1 as in scheme I proposed in Ref.~\cite{guimaraes2023noise}, hence simulations of Lindblad dissipators with negative decoherence rates are possible. A paradigmatic example of this is, for instance, the simulation of \emph{eternal non-markovianity} \cite{megier2017eternal}, which was numerically implemented in Section~\ref{sec: eternal_non_markovianity}.

Finally, we remark that we do not expect an increase of the number of circuits by amplifying the noise. Since we implement a quantum trajectory approach to increase the noise in the quantum simulation, the Pauli string sampling can be done simultaneously with the noise mitigation sampling required by PEC at the quantum circuit level. We note that the latter has an exponential cost as a function of the circuit volume, hence it remains as the main bottleneck of the quantum simulation.

The absence of the Trotter time-step adjustment (which is present in scheme I) enables one to potentially obtain relatively reduced mitigation factors in scheme II. This results in a reduction of computational resources comparatively to scheme I following Eq.~\eqref{iteration_mit_cost_local}. Consider a simulation where, for a given $\Delta t$, some initial decoherence rates are smaller (larger) than the target ones, namely, $\Gamma_{k}<\gamma_{k}$ ($\Gamma_{k}>\gamma_{k}$). For those which are smaller, the mitigation factors obtained with the proposed decoherence rate control scheme II are:
\begin{equation}
    r_{k}=1-\frac{\Delta \Gamma_{k}\Delta t}{\epsilon_{k}} . \label{mitigation_factor_scheme_II}
\end{equation}
In contrast to the definition of the mitigation factor of scheme I in Eq.~\eqref{mitigation_factor_scheme_I}, each mitigation factor $r_{k}$ is independent of other decoherence rates  $\Gamma_{k'\neq k}$. 

\subsection{Schemes' performance analysis}

To illustrate some of the advantages of scheme II over scheme I, we consider a quantum circuit, for simplicity, with uniform characterized error probabilities $\epsilon_{k}=\epsilon_{mean}$ and random target error probabilities, $\epsilon_{k}$, sampled from a Gaussian probability distribution, with a mean $\epsilon_{mean}$ and a standard deviation $\sigma$.  This sampling procedure is summarized in Fig.~\ref{fig:scheme_analysis}(a). Quantum circuits implemented in real quantum hardware typically contain highly non-uniform (i.e. disperse) error probabilities \cite{chen2023learnability,van2023probabilistic,guimaraes2023noise}. We gradually increase the standard deviation $\sigma$ of the Gaussian distribution, such that the target error probabilities become increasingly more disperse and compute the averaged total mitigated error probability, $\overline{\epsilon_{r}}$, obtained by each scheme. We note that $\overline{\epsilon_{r}}$ has a large influence on how many circuits one needs to execute in order to simulate the target decoherence rates, which in the worst case is $\exp{2\lambda (n-1) \overline{\epsilon_{r}}}$ for each Trotter iteration (see Eq.~\eqref{cost_scaling2}). In Fig.~\ref{fig:scheme_analysis}(b), we plot the results obtained. We observe that $\overline{\epsilon_{r}}$ for scheme II is consistently smaller than the one of scheme I when the target decoherence rates become more disperse (i.e. larger standard deviation $\sigma$). This suggests that scheme II performs better as the desired target decoherence rates and error probabilities in the quantum circuit become more disperse. To quantify how better the performance of scheme II is relatively to scheme I, in Fig.~\ref{fig:scheme_analysis}(c), we present the calculated average ratio $R_{II,I}=C_{\rm iter}^{\rm (II)}/C_{\rm iter}^{\rm (I)}$ for different numbers of qubits and $\epsilon_{mean}$, where $C_{\rm iter}^{\rm (I)}$ ($C_{\rm iter}^{\rm (II)}$) is the circuit-blind iteration mitigation cost for scheme I (II). Note that this ratio reflects the relative performance of each scheme for different numbers of qubits and target decoherence rates. For target decoherence rates sampled from a Gaussian distribution with a standard deviation shown by the gray vertical line in Fig.~\ref{fig:scheme_analysis}(b), we observe a linear reduction of the average ratio $R_{II/I}$ shown in Fig.~\ref{fig:scheme_analysis}(c). We also observe that the larger the mean error probability $\epsilon_{mean}$ of each stochastic Pauli channel is, the larger is the reduction of $R_{II,I}$. For instance, for $40$ qubits and $\epsilon_{mean}=0.04$, the reduction of $C_{\rm iter}^{\rm (II)}$ (scheme II) relatively to $C_{\rm iter}^{\rm (I)}$ (scheme I) is about $20 \%$. We note that this saving can be dramatic for deep circuits ($D>>1$), since the required number of circuits to be executed with scheme II scales as $(R_{II,I}C_{\rm iter}^{\rm (I)})^{2D}$. For instance, for $D=15$ Trotter layers in the previous example, the average reduction of computational resources, i.e. average number of circuits to be executed, relatively to scheme I already amounts to a factor of $10^{3}$. 

\begin{figure}
\centering
\includegraphics[width=0.4\textwidth]{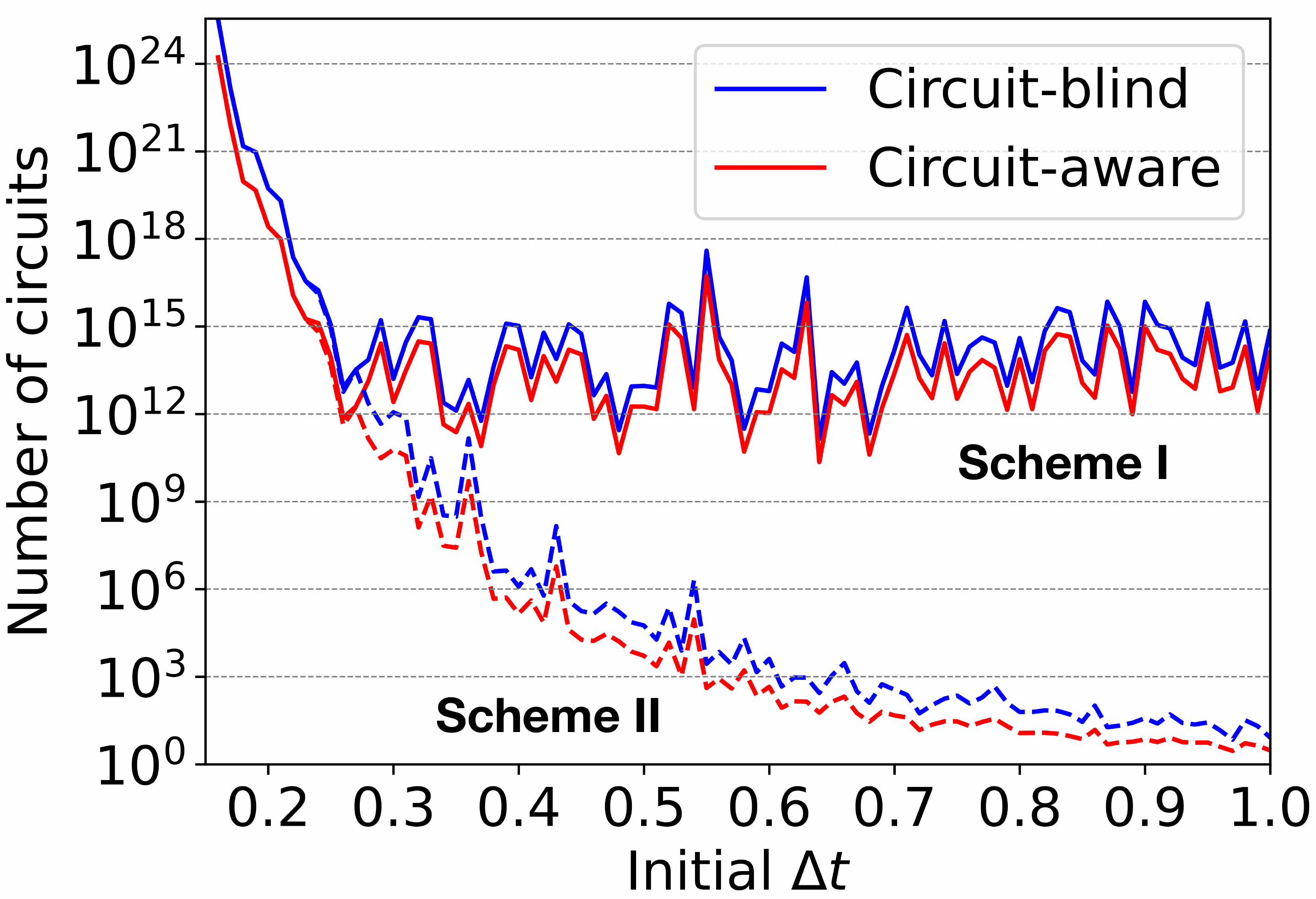}
\caption{Scaling of the number of circuits required for error mitigation using scheme I (II) as solid (dashed) lines with a circuit-blind (circuit-aware) mitigation protocols in blue (red) as a function of the initial Trotter time-step $\Delta t$. We considered $n=20$ qubits encoded in the quantum circuit, $\lambda=0.5$, $k=1$-qubit observable, total simulation time $t=10$ (arb. units) and the total mitigation cost was computed by averaging over $500$ samples with fixed characterized error probabilities $\epsilon_{k}=0.05$. The target decoherence rates were sampled from a Gaussian distribution with mean $\Gamma^{\rm (mean)}_{k}=0.05$ and standard deviation $\sigma_{k}=0.025$ (negative decoherence rates were reset to $0$). The total mitigation cost is averaged over 500 samples for each $\Delta t$.}
\label{fig:delta_t_analysis}
\end{figure}

With scheme II, one can choose an arbitrary large Trotter time-step that minimizes the number of implemented Trotter iterations, i.e. minimizes circuit depth, and decreases the mitigation factor per Eq.~\eqref{mitigation_factor_scheme_II} relatively to scheme I. Therefore, a larger Trotter time-step (relatively to $\Delta t_{\rm max}$ of scheme I) in scheme II enables a two-fold reduction effect of the mitigation cost following Eq.~\eqref{local_mitigation_cost}, i.e. by decreasing the number $D$ of Trotter layers and the averaged total mitigated error probability $\overline{\epsilon_r}$, at the cost of a larger Trotter decomposition error. This is illustrated in Fig.~\ref{fig:delta_t_analysis}, where we calculate the average number of circuits required for circuit-aware and circuit-blind mitigation with schemes I and II, i.e. $(C_{tot}^{\rm (II)})^2$ as a function of the initial $\Delta t$ for uniform characterized error rates $\epsilon_{k}=0.02$ and Gaussian distributed target decoherence rates with mean $\Gamma^{\rm (mean)}_{k} = 0.05$ and standard deviation $\sigma_{k}=\Gamma^{\rm (mean)}_{k}/2$. We observe that scheme II requires a smaller number of circuits than scheme I to be executed as a function of the initial Trotter time-step for initial $\Delta t > \Delta t_{max} \sim 0.25$. For small $\Delta t < \Delta t_{max}$, i.e. when initial decoherence rates are all above the target ones, both schemes require only error mitigation (in scheme I, this means only step 2 is required), therefore both schemes perform equivalently. However, as $\Delta t$ surpasses $\Delta t_{max}$, step 1 of scheme $I$ requires the change of the initial $\Delta t$ to $\Delta t_{max}$, hence all decoherence rates are increased. Some of these must be mitigated in step 2, namely, with an increased mitigation factor than in scheme II, which does not require change of the initial $\Delta t$. The break-even point defined by $\Delta t_{max}$ becomes smaller as mean target decoherence rates increase as can be perceived by Eq.~\eqref{delta_t_max}, i.e. $\Delta t_{max} \sim \left(\Gamma^{\rm (mean)}_{k}\right)^{-1}$. Therefore, we expect scheme II to vastly outperform scheme I for medium and large target decoherence rates $\Gamma^{\rm (mean)}_{k}$. These results illustrate the critical influence of being able to choose an arbitrary $\Delta t$ in the simulation. They also show that the circuit-aware mitigation protocol can reduce the computational costs relative to the circuit-blind protocol using our proposed decoherence rate control scheme II. Following Eq.~\eqref{local_mitigation_cost}, we expect that the larger the size of the system is, the larger will be the gap of the required number of circuits to be executed in the quantum device between the circuit-blind and circuit-aware mitigation protocols.

These results also suggest that scheme II is, in general, more suitable to be employed than scheme I. For disperse, high error rates or large numbers of qubits in the quantum circuit, i.e. the expected regime of large-scale quantum simulations in current quantum devices, scheme II is better suited for the task of controlling the decoherence rates than scheme I. 
\section{Quantum simulation of perturbative non-Markovian dynamics}\label{sec: simulation_time_dependent_theory}
In this section, we briefly review the Lindblad equation with time-dependent coefficients and introduce a protocol to simulate it via a noise-assisted digital quantum simulation on a noisy quantum processor. Such equation allows us to simulate perturbative non-Markovian dynamics on a noisy quantum processor without the requirement of ancilla qubits or mid-circuit measurements.
\subsection{Lindblad equation with time-dependent coefficients}
The ultimate aim of open dynamics is describing the evolution of the reduced density matrix of a system that interacts with a larger environment. Consider the system $S$ described by the Hilbert space $\mathcal{H}_S$ that interacts with an environment with Hilbert space $\mathcal{H}_E$. Solving the system-environment closed dynamics yields the unitary evolution operator $\hat{U}_{\rm SE}(t, 0)$, and the reduced density matrix of the system can be obtained by  $\hat{\rho} (t) = \Tr_E  [\hat{U}_{\rm SE}(t, 0) \hat{\rho}_{\rm SE}(0) \hat{U}_{\rm SE}^{\dagger} (t, 0)]$~\cite{rivas2012open}.  This complete operation can be seen as a map $\Lambda_t : \mathcal{B(H_S)} \longrightarrow \mathcal{B(H_S)}$, where $\mathcal{B(H_S)}$ is the set of bounded linear operators acting on $\mathcal{H}_{S}$. Unsurprisingly, solving this equation is difficult in practice, as it requires full description of the system and environment. In the same fashion as the Hamiltonian is the generator of closed dynamics, one can look for the generator of open system dynamics. The most general linear map for $\dot{\hat{\rho}}$ that preserves the trace and hermiticity of $\hat{\rho}_{\rm SE}$, and is local in time is~\cite{goriniCompletelyPositiveDynamical1976} ($\hbar=1$):
 \begin{align}\label{canonical_gksl}
         \mathcal{L}(t) [\hat{\rho}(t)] = & -i [\hat{H}(t),\hat{\rho}(t)]\\
            & + \sum_{k} \Gamma_k (t)   \biggl[ \hat{V}_k (t) \hat{\rho} (t) \hat{V}_k^\dagger (t) \nonumber \\   
          & -  \frac{1}{2} \left\{\hat{V}_k^\dagger(t) \hat{V}_k (t), \hat{\rho} (t) \right\} \biggr] ,
    \end{align} 
with $\partial \hat{\rho}(t)/\partial t =\mathcal{L}(t) [\hat{\rho}(t)]$, $\hat{H}(t)$ Hermitian, $\{\hat{V}_k (t)\}$ a set of Hermitian traceless orthogonal operators, and $\{\Gamma_k (t)\}$ scalar functions of time. This is the so-called canonical Gorini–Kossakowski–Sudarshan–Lindblad equation~\cite{hallCanonicalFormMaster2014}, which uniquely characterizes the dynamics. On the remaining of the manuscript, we will be using the Lindblad equation with time-dependent coefficients as a natural generalization of the time-independent equation introduced in Eq.~\eqref{Lindblad_eq}. As we will see, this natural extension allows for the simulation of non-Markovian dynamics \cite{goriniCompletelyPositiveDynamical1976, lindbladGeneratorsQuantumDynamical1976}. For more details about the properties of this equation, we refer the more interested reader to \cite{rivas2012open,chruscinskiDynamicalMapsMarkovian2022}.

\subsection{Time-dependent noise-assisted digital quantum simulation} \label{sec:time-dependentLindblad}

A noise-assisted digital quantum simulation \cite{guimaraes2023noise} may also be generalized to simulate time-dependent coefficients in Eq.~\eqref{canonical_gksl}. In what follows, we discuss the implementation of the decoherence rate control scheme II introduced in Sec.~\ref{sec: scheme II} to such open system models (with time-independent Hamiltonians).

\subsubsection{Stochastic Pauli noise}

Using time-dependent error mitigation (see Sec.~\ref{sec: timedependent_PEC}) and the decoherence rate control scheme II (see Sec.~\ref{sec: scheme II}), the decoherence rate in the quantum circuit may be adjusted either by independently mitigating or amplifying the noise in each Trotter layer.

Interestingly, using this method, the target decoherence rates can also be taken to be temporarily negative by choosing $r_{k} >1$ (see Eq.~\eqref{eq: nonMarkovian_discretized_gamma}). This allows one to simulate negative decoherence rates at the cost of a higher sampling cost. This is, with $r_{k}>1$ the total mitigation cost is increased relatively to a full error-mitigated quantum simulation with $r_{k} = 1$ as can be perceived by inspecting Eq.~\eqref{cost_scaling2}.

Open systems with stochastic Pauli noise dissipators appear in the modelling of some realistic systems, for instance, on the damping of exciton Rabi rotations by acoustic phonons in driven quantum dots \cite{ramsayDampingExcitonRabi2010}.

\begin{figure*}
\centering
\includegraphics[width=\textwidth]{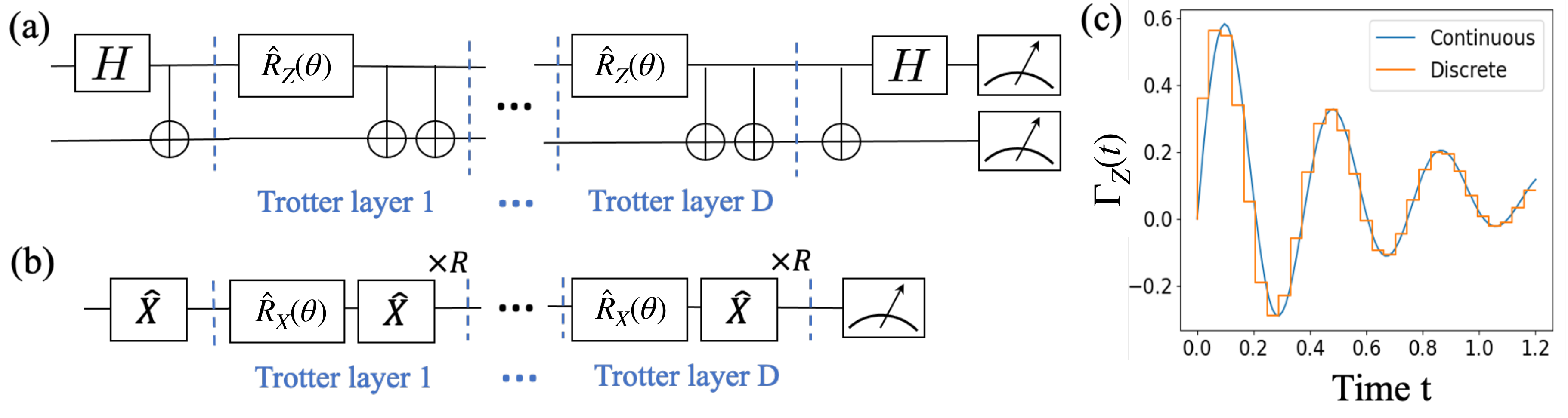}
\caption{(a) 2-qubit circuit implemented to simulate the Lindblad equation shown in Eq.~\eqref{oscillating_decoherence_rates}. Before (After) the Trotter layers we implement a Bell state initialization (measurement). (b) 1-qubit circuit used to simulate eternal non-Markovianity and time-dependent amplitude damping rates. The $\hat{X}$ operators are introduced to introduce single-qubit Pauli noise in each Trotter layer, therefore $R$ is always even. (c) Discrete and their continuous version of the time-dependent decoherence rates that were used in the quantum simulation in Section~\ref{sec: oscillating_decoherence_rates}. The parameters used in the simulations were $\theta = 2 E \Delta t$ and $R=30$.}
\label{fig:circuits_gamma}
\end{figure*}

\subsubsection{Amplitude damping}

The simulation of time-dependent single-qubit (generalized) amplitude damping can also be performed via a noise-assisted technique without the requirement of ancilla qubits or mid-circuit measurements \cite{guimaraes2023noise}. This only requires the ability to perform high-fidelity reset operations in the quantum device. Such high-quality operations have been recently achieved, for instance, in superconducting and ion-trap quantum platforms \cite{mcewen2021removing,mi2023stable,chertkov2022holographic}.

In view of simulating time-dependent amplitude damping, we consider the implementation of the following generalized reset channel to a quantum circuit,
\begin{align}
    \mathcal{R}(t)[\hat{\rho}] &= [1-p(t)]\hat{I}\hat{\rho}\hat{I}+p(t)\mathcal{V}(\hat{\rho}),\label{generalized_channel}\\
    \mathcal{V}(\hat{\rho}) &= \ket{\Psi}\bra{\Phi} \hat{\rho} \ket{\Phi}\bra{\Psi} + \ket{\Psi}\langle\Phi^{\perp}|\hat{\rho} |\Phi^{\perp}\rangle\bra{\Psi},
\end{align}
where $\ket{\Psi}=\hat{V}\ket{0}_{m}$, $\ket{\Phi}=\hat{U}\ket{0}_m$ and $|\Phi^{\perp}\rangle=\hat{U}\ket{1}_m$ with $\hat{U}^{\dag}$ and $\hat{V}$ denoting single-qubit gates applied before and after the reset operation acting on qubit $m$. The time-dependent stochastic application of the generalized reset operation to each Trotter layer with discretized probability $p(t=D\Delta t)$ leads to a Lindblad dissipator
\begin{equation}
    \mathcal{D}_{\rm AD}(t)[\hat{\rho}] = \mathcal{D}_{\ket{\Psi}\bra{\Phi}}(t)[\hat{\rho}]+\mathcal{D}_{\ket{\Psi}\bra{\Phi^{\perp}}}(t)[\hat{\rho}],
\end{equation}
where
\begin{align}
\mathcal{D}_{\ket{\alpha}\bra{\beta}}(t)[\hat{\rho}]&=\gamma(t)\left(\ket{\alpha}\bra{\beta} \hat{\rho} \ket{\beta}\bra{\alpha}
    -\frac{1}{2}\{ \ket{\beta}\bra{\beta},\hat{\rho}\}\right), \\
    \gamma(t) &= p(t)/\Delta t. \label{eq:gamma_ad}
\end{align}
Let us choose, for simplicity, $\hat{U} = \hat{V}$. Then, the generalized reset channel leads to the time-dependent amplitude damping and local dephasing in the $\{\ket{\Phi},|\Phi^{\perp}\rangle\}$ basis, expressed as,
\begin{align}
    \mathcal{D}_{\rm AD}(t)[\hat{\rho}] &= 
    \gamma(t) \ket{\Phi}\langle \Phi^{\perp}|\hat{\rho} |\Phi^{\perp}\rangle\bra{\Phi}- \frac{\gamma(t)}{2}\left\{|\Phi^{\perp}\rangle\langle \Phi^{\perp}|,\hat{\rho}\right\} \label{eq: AD_definition}\\
    &\quad + \frac{\gamma(t)}{4}\left( \hat{U}\hat{Z}_{m}\hat{U}^{\dagger}\hat{\rho} \hat{U}\hat{Z}_{m}\hat{U}^{\dagger} - \hat{\rho} \right)\nonumber, 
\end{align}
where $\hat{U}\hat{Z}_{m}\hat{U}^{\dagger}=\ket{\Phi}\bra{\Phi}-|\Phi^{\perp}\rangle\langle \Phi^{\perp}|$. Therefore, by taking $\hat{U}\hat{Z}_{m}\hat{U}^{\dagger}$ to be a Pauli operator, such as $\hat{X}_m$, one can implement time-dependent amplitude damping in the eigenbasis of the Pauli operator of choice, while the additional dephasing rate $\gamma(t)/4$ is controlled by our time-dependent partial noise mitigation scheme, introduced in Sec.~\ref{sec: decoherence_schemes} and Sec.~\ref{sec:time-dependentLindblad}. One can introduce different combinations of non-unital noise channels with a sequential application of different generalized reset channels. One may implement, for instance, relaxation noise \cite{guimaraes2023noise}, also known as generalized amplitude damping \cite{nielsen2002quantum} to simulate realistic open systems such as exciton-phonon interactions in quantum dots~\cite{nazirModellingExcitonphononInteractions2016}.

\begin{figure*}
\centering
\includegraphics[width=\textwidth]{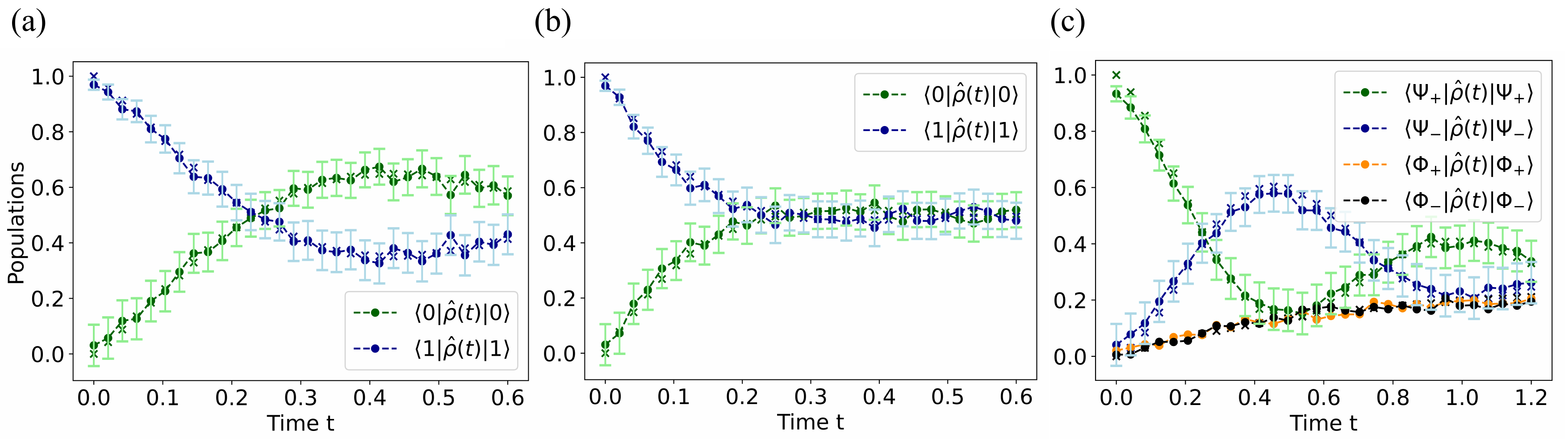}
\caption{Results of the digital quantum simulations performed with controlled noise (dots) and classically solved Lindblad equations (crosses). (a) Eternal non-Markovianity results. (b) Same open system as in (a) but with time-independent Lindblad dissipators as shown in Eq.~\eqref{time-independent_Lindblad_dissipator}. (c) Lindblad dissipator with oscillating decoherence rate as per Eq.~\eqref{oscillating_decoherence_rates}. We executed $180(C_{tot}^{\rm (loc)})^2$ (see Eq.~\eqref{cost_scaling2}) circuits for each data point with $E=\pi$ (arb. units).}
\label{fig:results_evol}
\end{figure*}

\section{Numerical implementation}\label{sec: num_impl}

We implement the decoherence rate control scheme II described in the Section~\ref{sec: scheme II} on the emulated IBM Quantum device \emph{ibmq mumbai} to simulate several Lindblad models with time-dependent coefficients. We first characterize the stochastic Pauli noise in a Trotter layer as described in Section~\ref{time_evol_sec} and then perform the noise-assisted digital quantum simulation as discussed in Section~\ref{sec:time-dependentLindblad}.

\subsection{Eternal non-Markovianity}\label{sec: eternal_non_markovianity}

Here we simulate an 1-qubit with Hamiltonian $\hat{H}=E \hat{X}$ (where $E=$~const), which is acted upon with a Lindblad dissipator given as follows:
\begin{align}
        \mathcal{D}_{\rm eternal}(t)[\hat{\rho}]=& \left(\hat{X}\hat{\rho}\hat{X}-\hat{\rho}\right)+\left(\hat{Y}\hat{\rho}\hat{Y}-\hat{\rho}\right) \nonumber \\
        & -\tanh{(t)}\left(\hat{Z}\hat{\rho}\hat{Z}-\hat{\rho}\right).\label{eternal_non_markovianity_lindblad_dissipator}
\end{align}
This is one of the characteristic cases of non-Markovianity, so-called \emph{eternal non-Markovianity} \cite{megier2017eternal}, where one of the decoherence rates is negative for times $t>0$. We initialize the qubit in the state $\ket{1}$, apply the Trotter formula to evolve the system and finally measure the population terms of the density matrix as illustrated in Fig.~\ref{fig:circuits_gamma}(b). Within each Trotter layer, we apply an even number $R$ of Pauli operators $\hat{X}$ in sequence to introduce single-qubit Pauli errors in the circuit (Randomized Compiling was applied to the layer of $R$ $\hat{X}$ gates in each Trotter layer). Note that the ideal version of these gates cancel each other for even $R$, hence the logical circuit remains the same. However, the waiting time required to implement all $R$ $\hat{X}$ gates introduces noise in the quantum circuit in the form of amplitude and phase damping, where the former is transformed into stochastic Pauli noise via RC.

The quantum simulation results are shown in Fig.~\ref{fig:results_evol}(a). We also solved a Lindblad equation with time-dependent coefficients in a classical computer using a Trotter formula approach. We find that the quantum and classical simulation results are similar. For sake of comparison, in Fig.~\ref{fig:results_evol}(b), we plot the results of same open system model but with time-independent noise instead, defined as,
\begin{equation}
        \mathcal{D}_{\rm comp}[\hat{\rho}]=2\left[\left(\hat{X}\hat{\rho}\hat{X}-\hat{\rho}\right)+\left(\hat{Y}\hat{\rho}\hat{Y}-\hat{\rho}\right)+\left(\hat{Z}\hat{\rho}\hat{Z}-\hat{\rho}\right)\right]\label{time-independent_Lindblad_dissipator}.
\end{equation}
The larger decoherence rates in Eq.~\eqref{time-independent_Lindblad_dissipator} relatively to Eq.~\eqref{eternal_non_markovianity_lindblad_dissipator} yield a faster relaxation to the steady-state as can be seen in Fig.~\ref{fig:results_evol}(b) relatively to Fig.~\ref{fig:results_evol}(a).

\subsection{Oscillating Pauli decoherence rate}\label{sec: oscillating_decoherence_rates}
We initialize a 2-qubit system in a Bell state $\ket{\Psi_{+}}=\frac{1}{\sqrt{2}}\ket{00}+\ket{11}$, evolve the first qubit with the Hamiltonian $\hat{H}=E \hat{Z}$ and measure the populations of the Bell states. On each Trotter layer, we add $2$ CNOT gates in order to introduce non-negligible noise to each Trotter layer (Randomized Compiling was applied to each pair of CNOT gates). Note that the logical circuit remains the same because the CNOT gates cancel each other, however due to their faulty implementation, noise is introduced in the circuit. In Fig.~\ref{fig:circuits_gamma}(a), we illustrate the quantum circuit implemented on the emulated quantum device.

We considered the following Lindblad dissipator to be simulated \cite{svozilik2020universal}:
\begin{align}
    \mathcal{D}_{\rm osc}(t)[\hat{\rho}]&=\Gamma_{Z}(t)\sum_{m=1}^{2}(\hat{Z}_m \hat{\rho}\hat{Z}_m - \hat{\rho}) \label{oscillating_decoherence_rates}\\
    &\quad+ \Gamma_{X,Y}\sum_{m=1}^{2}(\hat{X}_{m} \hat{\rho}\hat{X}_{m} +\hat{Y}_{m} \hat{\rho}\hat{Y}_{m}- 2\hat{\rho}),\nonumber
\end{align}
where we considered time-dependent decoherence rates $\Gamma_{Z}(t)$ for the single-qubit dephasing term $\hat{Z}$ acting on each qubit and time-independent decoherence rates $\Gamma_{X,Y}=0.2$ associated to single-qubit Pauli interactions $\hat{X}$ and $\hat{Y}$ acting on each qubit. The time-dependent coefficients $\Gamma_{Z}(t)$ follow a Jaynes-Cummings model for a system of two qubits where each one interacts with a detuned optical cavity characterized by a Lorentzian spectral function~\cite{svozilik2020universal}. Additionally, we added, for sake of demonstration, Lindblad dissipators with time-independent coefficients as shown in the last line of Eq.~\eqref{oscillating_decoherence_rates}. All other remaining 2-qubit dissipative Pauli interactions in the quantum circuit have been fully mitigated. We discretized $\Gamma_{Z}(t)$ as shown in Fig.~\ref{fig:circuits_gamma}(c).

In Fig.~\ref{fig:results_evol}(c), we show the results of measuring the population terms of the density matrix in the Bell basis for the quantum and classical simulations. We observe that the quantum simulation results follow closely the classically solved ones.

\begin{figure}
\centering
\includegraphics[width=0.48\textwidth]{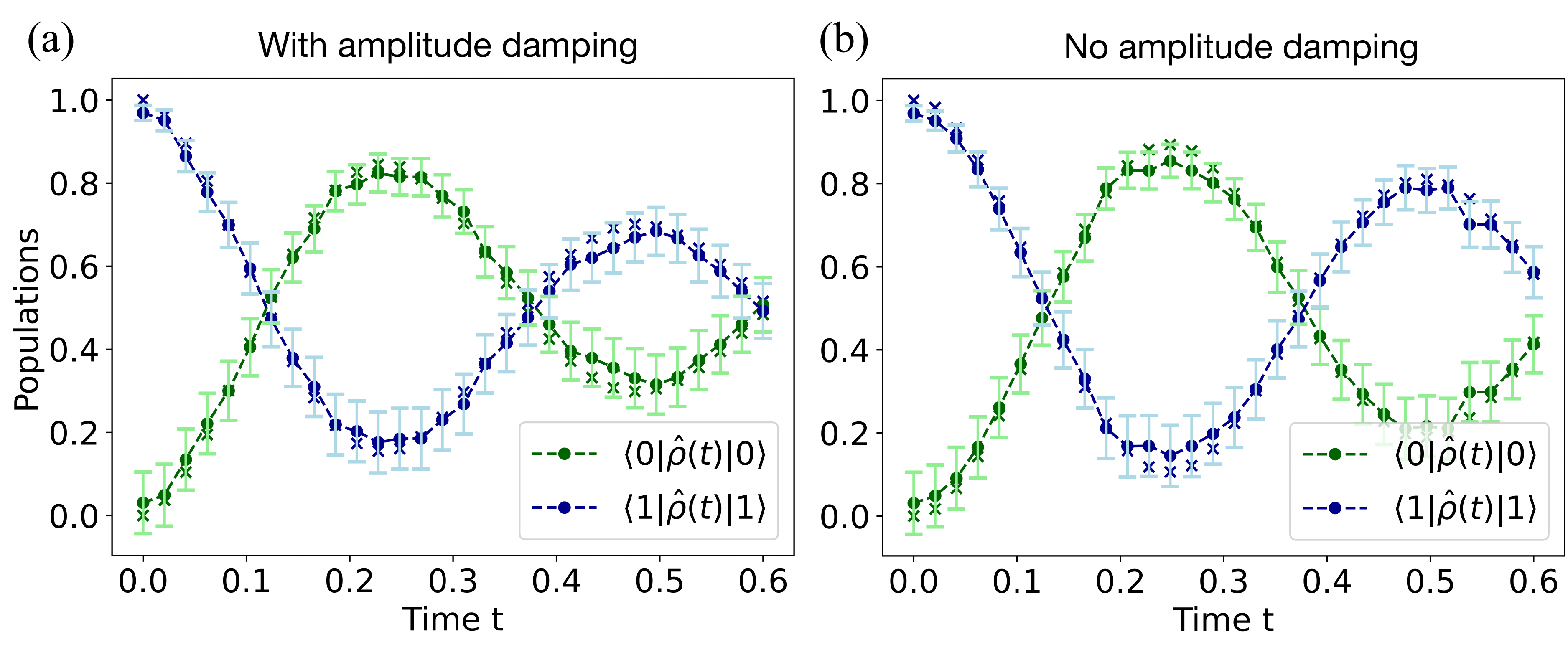}
\caption{Results of the digital quantum simulations performed with controlled noise (dots) and classically solved Lindblad equations (crosses). (a) Time-dependent amplitude damping ($\Gamma_{\rm AD}=1+\tanh(t)$) and time-independent dephasing ($\Gamma_{Z}=1$) Lindblad dissipators. (b) Time-independent dephasing Lindblad dissipator with $\Gamma_{\rm AD}=0$ and $\Gamma_{Z}=1$. We executed $180(C_{tot}^{\rm (loc)})^2$ (see Eq.~\eqref{cost_scaling2}) circuits for each data point with $E=2.1\pi$ (arb. units).}
\label{fig:AD}
\end{figure}

\subsection{Time-dependent amplitude damping rate}\label{sec: amp_damp}
We implement the same 1-qubit circuit used in the Sec.~\ref{sec: eternal_non_markovianity} and illustrated in Fig.~\ref{fig:circuits_gamma}(b). We renormalize the Hamiltonian to $\hat{H}=E\hat{X}$, and implement  the following phenomenological Lindblad dissipator,
\begin{align}
    \mathcal{D}_{\rm AD}^{\rm (num)}(t)[\hat{\rho}]&=\Gamma_{\rm AD}(t)\left(\ket{0}\langle 1|\hat{\rho} |1\rangle\bra{0}- \frac{1}{2}\left\{|0\rangle\langle 1|,\hat{\rho}\right\} \right) \nonumber \\
    & + \Gamma_{Z}(\hat{Z} \hat{\rho}\hat{Z} - \hat{\rho})\label{eq: num_impl_AD},
\end{align}
where $\Gamma_{Z}=1$ denotes a time-independent decoherence rate and $\Gamma_{\rm AD}(t)=1+\tanh(t)$. We control the dephasing noise arising from $\hat{Z}$ and fully mitigate the remaining Pauli dissipative interactions, namely $\hat{X}$ and $\hat{Y}$. In order to simulate time-dependent amplitude damping, we chose $\hat{U}=\hat{V}=\hat{I}$ in Eq.~\eqref{eq: AD_definition} and we controlled via our decoherence rate control scheme II the extra $\hat{Z}$ Pauli string that arises from applying the generalized reset channel, as represented by the last term in Eq.~\eqref{eq: AD_definition}. We considered the application of the reset operation to last about $250$~$ns$, similar to 2-qubit gates and a failure reset channel $\mathcal{E}_{\rm er} = p_{\rm er}\hat{I}\hat{\rho}\hat{I}+(1-p_{\rm er})\mathcal{R}(t)[\hat{\rho}]$, being $p_{\rm er}\sim 10^{-3}$ the reset failure probability, in accordance to previous reports \cite{mcewen2021removing,mi2023stable}.

In Fig.~\ref{fig:AD}(a) and Fig.~\ref{fig:AD}(b), we plot the quantum and classical simulation outcomes of measuring the population terms of the density matrix with and without amplitude damping ($\Gamma_{\rm AD}(t)=0$), respectively. As expected, we observe that oscillations in the population terms show a smaller amplitude with the time-dependent amplitude damping dissipator in Fig.~\ref{fig:AD}(a) than without it as shown in Fig.~\ref{fig:AD}(b).
We also remark that the reset failure probability has an error probability of about one order of magnitude smaller than the ones we are simulating, hence the simulation's accuracy is not compromised. This reset failure probability can also be mitigated by adjusting the probabilities $p(t)$ in Eq.~\eqref{generalized_channel}.

\vspace{4mm}

\section{Conclusion}
In this manuscript, we have advanced the field of noise-assisted digital quantum simulations \cite{guimaraes2023noise}, specifically targeting large-scale open quantum systems. Our contributions are two-fold. First, we have integrated a circuit-aware mitigation protocol, as delineated in \cite{tran2023locality}, into the foundational noise-assisted quantum simulation algorithm. This integration results in a dramatic reduction in sampling cost compared to traditional, circuit-blind partial error mitigation strategies. Second, we introduce a novel decoherence rate control scheme that further minimizes sampling requirements, especially in scenarios involving high error rates and larger systems—areas that are pertinent for large-scale open system simulations.

Furthermore, we extend the noise-assisted simulation algorithm to include Lindblad equations with time-dependent coefficients. By incorporating time-dependent mitigation factors alongside our newly proposed decoherence rate control scheme, we enable the simulation of time-dependent (positive and negative) decoherence rates, thereby capturing non-Markovian dynamics. We validate these enhancements through simulations involving 1- and 2-qubit non-Markovian systems.


Looking ahead, several intriguing avenues for future research emerge. One immediate goal is the experimental validation of our optimized noise-assisted simulation algorithm on actual quantum hardware. Constraints like extended device access times have prevented real-device implementation in this study. Moreover, the algorithm's adaptability to different quantum computing paradigms, such as ion-trap systems with full qubit connectivity, could offer additional opportunities for optimization. Lastly, extending the algorithm to solve other dynamical equations like the Redfield or Hierarchical Equations of Motion could further broaden its applicability.

\section{Acknowledgements}
JDG and ARM acknowledge funding from the 
Portuguese Foundation for Science and Technology (FCT) through PhD grants UI/BD/151173/2021 and SFRH/BD/151453/2021, respectively. JDG, JL, MBP acknowledge support by the BMBF 
project PhoQuant (grant no. 13N16110). MIV acknowledges support from the FCT through Strategic Funding UIDB/04650/2020. SFH and MBP acknowledge
support by the DFG via the QuantERA project ExtraQt. The authors acknowledge support by the state of Baden-Württemberg through bwHPC and the 
German Research Foundation (DFG) through Grant No. INST 40/575-1 FUGG (JUSTUS 2 cluster).

\bibliography{apssamp}

\end{document}